\documentclass[12pt]{article}
\usepackage{amssymb,graphicx,color,epsfig}
\begin{document}

\baselineskip=.22in
\renewcommand{\baselinestretch}{1.2}
\renewcommand{\theequation}{\thesection.\arabic{equation}}
\newcommand{\klmt}{\mbox{K\hspace{-7.6pt}KLM\hspace{-9.35pt}MT}\ }
\begin{flushright}
{\tt hep-th/0612285\\
KIAS-P06067}
\end{flushright}


\begin{center}
{{\Large \bf Cosmic D- and DF-strings from D${\bf 3{\bar D}3}$: \\[2mm]
Black Strings and BPS Limit}\\[12mm]
Taekyung Kim~~and~~Yoonbai Kim\\[2mm]
{\it BK21 Physics Research Division and Institute of Basic Science,}\\
{\it Sungkyunkwan University, Suwon 440-746, Korea}\\
{\tt pojawd@skku.edu, yoonbai@skku.edu}\\[5mm]
Bumseok Kyae\\[2mm]
{\it School of Physics, Korea Institute for Advanced Study,\\
207-43, Cheongryangri-Dong, Dongdaemun-Gu, Seoul 130-012, Korea}\\
{\tt bkyae@kias.re.kr}\\[5mm]
Jungjai Lee\\[2mm]
{\it Department of Physics, Daejin University,
 Pocheon, Gyeonggi 487-711, Korea}\\
{\tt jjlee@daejin.ac.kr} }
\end{center}

\vspace{5mm}

\begin{abstract}
We study D- and DF-strings in a D3${\bar {\rm D}}3$ system by using
Dirac-Born-Infeld type action. In the presence of an electric flux
from the transverse direction, we discuss gravitating thick D-string
solutions of a spatial manifold, ${\rm S}^{2}\times {\rm R}^{1}$, in
which straight D-strings stretched along the R${}^{1}$ direction are
attached to the south and north poles of the two-sphere. There is a
horizon along its equator, which means the structure of black
strings is formed. We also discuss the BPS limit for thin parallel
D- and DF-strings in both flat and curved spacetime. We obtain the
BPS sum rule for an arbitrarily-separated multi-string configuration
with a Gaussian type tachyon potential. At the site of each thin BPS
D(F)-string, the pressure takes a finite value. We find that there
exists a maximum deficit angle $\pi$ in the conical geometry induced
by thin BPS D- and DF-strings.
\end{abstract}


\newpage

\setcounter{equation}{0}
\section{Introduction}

Cosmic strings have been extensively studied in the context of
classical field theory, and Nielen-Olesen vortex-strings in the
Abelian-Higgs model have been regarded as the representative
example~\cite{VS}. By solving the nonlinear field equations in the
Abelian-Higgs model, one obtains point-like vortex solutions in
2-dimensional flat space. When such solitons gravitate in three
dimensional spacetime, their 1-dimensional stringy objects could be
interpreted as straight cosmic strings. The resulting conical
geometry with deficit angle~\cite{Vilenkin:1981zs} has cosmological
implications associated with possible lensing signatures, which are
significantly different from those of standard gravitational
lenses~\cite{Sazhin:2003cp}, or cosmological perturbations induced
by the motion of cosmic strings.

A characteristic property of some topological solitons, including
kinks, vortices, and monopoles, is saturation of the BPS
bound~\cite{MS}. In the case of the BPS vortices in the
Abelian-Higgs model, an analysis of moduli space dynamics based on
their $2n$ zero modes predicts 90 degree scattering in the head-on
collision of two slowly-moving vortices~\cite{Shellard:1988zx}. This
property has been confirmed by numerical analysis even for the
fast-moving vortices~\cite{Shellard:1988zx}, and it suggests an
inter-commuting process of two colliding strings, which plays an
important role in the evolution of a stringy network.

Another candidate for cosmic string is the
superstring~\cite{Witten:1985fp}. It shares almost all of the
characteristic with vortex-strings in gauge theories with
spontaneous symmetry breaking mechanism. As the topics of D-branes
and the warped geometry have been developed, the idea of the cosmic
superstring has recently been revived and, as an appropriate
candidate, the D- and DF-strings have attracted much
attention~\cite{Dvali:2003zj,Copeland:2003bj,Polchinski:2004ia}.
Concerning string cosmology~\cite{Cline:2006hu}, an intriguing
setting is D- and DF-strings generated from the system of D3${\bar
{\rm D}}3$, particularly those in the \klmt
model~\cite{Kachru:2003sx}.

For the generation of macroscopic D- and DF-strings from the system
of D3${\bar {\rm D}}3$, a useful description can be made by
employing the tachyon effective action~\cite{Sen:2003tm,Sen:2004nf}
as shown in the case of tachyon kinks~\cite{Kim:2003in}. When D3 and
${\bar {\rm D}}3$ are separated by a short distance, they start to
approach each other due to mutual attraction and finally collide.
Cosmologically, this period corresponds to an inflationary era in
the early universe~\cite{Dvali:1998pa}. Instability of D${\bar {\rm
D}}$ implies the presence of a complex tachyon field, and zero modes
on D and ${\bar {\rm D}}$ correspond to two gauge fields of the
U(1)$\times$U(1) symmetry. For description of this system, two kinds
of effective actions have been proposed: One is that derived from
boundary string field theory, which is very
complicated~\cite{Kraus:2000nj,Jones:2002si}, and the other is that
of a Dirac-Born-Infeld (DBI) type~\cite{Sen:2003tm,Garousi:2004rd}.
As a U(1) symmetry is spontaneously broken, codimension-two D- or
DF-branes are given as static vortex-string solutions in flat
spacetime. These are nothing but straight D- or DF-strings
(($p,q$)-strings) from the D3${\bar {\rm
D}}3$~\cite{Kim:2005tw,Cho:2005wp}. In order to detect fossils of
the cosmic superstrings, some remnants of macroscopic cosmic
superstrings should  survive in the present universe as far as their
result is consistent with the data from the cosmic microwave
background experiments. In this sense, D(F)-strings from D3${\bar
{\rm D}}3$ is deemed more viable since they are produced after
D-brane inflation~\cite{Sarangi:2002yt}.

Once cosmic superstrings are produced, various dynamical issues need
to be understood, e.g., collisions between D-strings, F-strings, and
DF-strings~\cite{Jackson:2004zg}, formations of
Y-junctions~\cite{Copeland:2003bj,Jackson:2004zg,Copeland:2006eh},
probability of the reconnection of cosmic
superstrings~\cite{Hanany:2005bc}, and evolution of a cosmic
DF-string network~\cite{Tye:2005fn}. When the D${\bar {\rm D}}$ is
dissociated, it decays into either perturbative closed string
degrees~\cite{Lambert:2003zr} or nonperturbative open string degrees
like cosmic D(F)-strings. In this sense, graviton production from D-
or DF-strings is intriguing~\cite{Cornou:2006wx}. Concerning string
cosmology based on flux compactification~\cite{Kachru:2003aw}, D-
and DF-strings in a warped geometry need also to be
studied~\cite{Firouzjahi:2006vp}.

In section 2, we introduce the DBI type action of D3${\bar {\rm
D}}3$ and investigate gravitating thick global D-strings with radial
electric flux. On the spatial manifold ${\rm S}^{2}\times {\rm
R}^1$, the straight strings, stretched along the ${\rm R}^1$, are
attached to both the south and north poles of the ${\rm S}^{2}$. A
horizon is formed along the equator of the ${\rm S}^{2}$. Thus, the
solution found is a black string
solution~\cite{Horowitz:1991cd,Kim:1997ba}. In section 3, through a
systematic study, we demonstrate the saturation of the BPS sum rule
in the thin-limit D- and DF-strings in both flat and curved
spacetime. We derive a condition for the tachyon potential
saturating the BPS sum rule of multi-D(F)-strings and show that
Gaussian type tachyon potential saturates this BPS sum rule,
consistent with the descent relation for codimension-two BPS branes.
The resulting deficit angle for thin BPS D- and DF-strings has the
maximum value of $\pi$. We conclude in section 4 with a summary of
the obtained results and brief discussions for further study.

\setcounter{equation}{0}
\section{Gravitating D-strings with Electric Flux}

A D3${\bar {\rm D}}$3 system coupled to gravity in their coincidence
limit can be described by the DBI type action with U(1)$\times$U(1)
gauge symmetry~\cite{Sen:2003tm,Garousi:2004rd}
\begin{eqnarray}
S&=&S_{{\rm EH}}+S_{{\rm M}}
\label{lact} \\
&=& \int d^{4}x \left\{ \frac{\sqrt{-g}}{2\kappa^{2}}(R-2\Lambda)
-2{\cal T}_{3}V(\tau) \left[\,\sqrt{-\det(X^{+}_{\mu\nu})}
+\sqrt{-\det(X^{-}_{\mu\nu})}\,\,\right] \right\}\, ,
\label{acX}
\end{eqnarray}
where
\begin{equation}\label{Xpm}
X^{\pm}_{\mu\nu}=g_{\mu\nu}+F_{\mu\nu}\pm C_{\mu\nu} +({\overline
{D_{\mu}T}}D_{\nu}T +{\overline {D_{\nu}T}}D_{\mu}T)/2.
\end{equation}
In the action (\ref{acX}), we have a complex tachyon field $T=\tau
e^{i\chi}$, two U(1) gauge fields $A_{\mu}$ and $C_{\mu}$ with their
field strength tensor
$F_{\mu\nu}=\partial_{\mu}A_{\nu}-\partial_{\nu}A_{\mu}$ and
$C_{\mu\nu}=\partial_{\mu}C_{\nu}-\partial_{\nu}C_{\mu}$, and the
covariant derivative of the tachyon field $D_{\mu}T=(\partial_{\mu}
-2iC_{\mu})T$.

Before beginning the detailed study, we briefly explain the
terminology of the various strings of our interest. The D- and
DF-strings stretched along the $z$-axis are distinguished by the
nonvanishing fundamental (F) string charge density along the
$z$-axis, $\Pi_z$. Since $\Pi_z$ is defined by the electric flux
density conjugate to the $z$-component of gauge field $A_z$,
$\Pi^z\equiv\frac{1}{2\sqrt{-g}}\frac{\delta S}{\delta\partial_t
A_z}=\frac{1}{\sqrt{-g}}\frac{\delta S}{\delta E_z}$, the
vortex-strings with the vanishing $E_z$ correspond to {\it
D-strings} and those with the nonvanishing $E_z$ to {\it DF-strings}
(or {\it (p,q)-strings}). Similarly to the vortex-strings in the
Abelian-Higgs model, the {\it global (local)} D- and DF-string
solutions are provided from the trivial (nontrivial) gauge field
$C_{\mu}$ of vanishing (nonvanishing) field strength tensor
$C_{\mu\nu}=0\,\,(C_{\mu\nu}\neq0)$.

For the analysis of D- and DF-string solutions with the cylindrical
symmetry, the monotonic connection of $V(\tau=0)=1$ and
$V(\tau=\infty)=0$ is enough~\cite{Sen:1999xm}. For numerical
analysis, we use a specific form of the potential satisfying all of
the above conditions
\begin{equation}\label{V3}
V(\tau)=\frac{1}{{\rm cosh}\left(\frac{\tau}{R}\right)}.
\end{equation}

Since we are interested in straight black D-strings, it is
convenient to use cylindrical coordinates $(t,r,\theta ,z)$ in
curved spacetime. To be specific we take into account the simplest
form with translational symmetry along the $z$-direction:
\begin{equation}\label{trme}
d s^2 = -N(r)^{2} dt^2 + dz^{2} + b(r)(dr^{2}+r^{2}d\theta^{2}).
\end{equation}
Under the metric (\ref{trme}), the complex tachyon field $T$ with
$n$ vortices superimposed at the origin has
\begin{equation}\label{ant}
T=\tau(r) e^{in\theta}.
\end{equation}
We assume the fundamental strings stretched along a transverse
direction at $r=0$. In the next section, we consider additionally
those superposed along the straight cosmic D-strings. Their string
charge densities couple to the following components of electric
fields, respectively:
\begin{equation}\label{ane}
F_{tr}(r)=E_{r}(r),\quad F_{tz}(r)=E_{z}(r).
\end{equation}
To change global strings of our interest to local strings, we can
turn on the angular component of the other U(1) gauge field
$C_{\mu}$ and the corresponding field strength tensor $C_{\mu\nu}$
as
\begin{eqnarray}\label{anc}
C_{\mu}=\delta_{\mu \theta}C_{\theta}(r), \qquad C_{r\theta}=C_{\theta}'.
\end{eqnarray}

Substitution of the metric and ans\"{a}tze,
(\ref{trme})--(\ref{anc}), leads to the equations of motion. The
equation of motion for the tachyon amplitude $\tau$ is
\begin{eqnarray}\label{ter}
\lefteqn{
\frac{d}{dr}\left\{ \frac{\sqrt{-g}V}{\sqrt{X}}\left[1+
\frac{\tau^2}{br^2}(n-2C_\theta)^2\right]\frac{\tau'}{b}
\left(1-\frac{E_{z}^2}{N^2}\right)\right\}
}\nonumber\\
&&-\frac{\sqrt{-g}V}{\sqrt{X}}\left[
\left(1-\frac{E_{z}^2}{N^2}\right) \left(1+\frac{\tau^{'2}}{b}\right)
-\frac{E_{r}^2}{N^2
b}\right]\frac{\tau(n-2C_\theta)^2}{br^{2}}
=\sqrt{-g}\sqrt{X}~\frac{dV}{d\tau} ,
\end{eqnarray}
and that for the gauge field $C_{\theta}$ is
\begin{eqnarray}\label{ger}
\frac{d}{dr}\left[ \frac{\sqrt{-g}V}{\sqrt{X}}\left(1-
\frac{E_z^2}{N^2}\right) \frac{{C_{\theta}'}^2}{b^2r^2}\right]
+2\frac{\sqrt{-g}V}{\sqrt{X}}\left[\left(1-\frac{E_{z}^2}{N^2}\right)
\left(1+\frac{\tau^{'2}}{b}\right)-\frac{E_{r}^2}{N^2
b}\right]\frac{\tau^2(n-2C_\theta)}{br^{2}}=0 ,
\end{eqnarray}
where $-g=N^{2}b^{2}r^{2}$ and the determinant $X$ is
\begin{eqnarray}\label{Xr}
X&=&\frac{-\det(X^{\pm}_{\mu\nu})}{-g}\nonumber\\
&=&\left[1+\frac{\tau^2}{br^2}(n-2C_{\theta})^2\right]\left[
\left(1-\frac{E_{z}^2}{N^2}\right) \left(1+\frac{{\tau'}^2}{b}\right)
-\frac{E_{r}^2}{N^2
b}\right]+\left(1-\frac{E_{z}^2}{N^2}\right)\frac{{C_{\theta}'}^2}{b^2 r^2}.
\end{eqnarray}
The Bianchi identity, $\partial_{\mu}F_{\nu\rho}
+\partial_{\nu}F_{\rho\mu}+\partial_{\rho}F_{\mu\nu}=0$, dictates
that $E_{z}$ in (\ref{ane}) should be constant, and the conjugate
momentum $\Pi^{z}$ of $A_{z}$ is given by
\begin{eqnarray}\label{piz}
\Pi^{z} \equiv\frac{1}{\sqrt{-g}}\frac{\delta S}{\delta
E_{z}}=\frac{2{\cal T}_3
V}{\sqrt{X}}\left\{\left[1+\frac{\tau^2}{br^2}(n-2C_{\theta})^2
\right]\left(1+\frac{\tau^{'2}}{b}\right)+\frac{C_{\theta}^{'2}}{b^2
r^2}\right\}\frac{E_z}{N^2}.
\end{eqnarray}
The equation for the gauge field $A_{\mu}$ reduces to
$(Nbr\Pi^{r})'=0$ which results in the expression of conjugate
momentum $\Pi^{r}$ of $A_{r}$,
\begin{equation}\label{pir}
\Pi^{r}\equiv\frac{1}{\sqrt{-g}}\frac{\delta S}{\delta
E_{r}}=\frac{Q}{Nbr},
\end{equation}
where $Q$ is the fundamental string charge per unit length along the
transverse direction. Therefore, $E_{r}(r)$ is given by an algebraic
equation
\begin{equation}\label{er}
\frac{E_{r}^2}{N^2b} = \frac{\left(1-\frac{E_z
^2}{N^2}\right)\left\{\frac{C_\theta ^{'2}}{b^2r^2} +
\left(1+\frac{\tau'^2}{b}\right)\left[ 1+\frac{\tau^2}{b
r^2}\left(n-2C_\theta\right)^2\right]\right\}}
{\left[1+\frac{\tau^2}{b
r^2}\left(n-2C_\theta\right)^2\right]\left\{1+b\left(\frac{2{\cal
{T}} _ {3} V}{N b\Pi^r}\right)^2\left[1+\frac{\tau^2}{b
r^2}\left(n-2C_\theta\right)^2\right]\right\}}.
\end{equation}
Since the fundamental strings of the charge $Q$ are not localized
along the vortex-strings but stretched to the transverse direction,
the D-string carrying the fundamental string charge $Q$ cannot be a
DF-string but just be a {\it D-string with electric flux.} Among
four nonvanishing $(t,r,\theta , z)$-components of the Einstein
equations, we choose two independent equations for the metric
functions, $N$ and $b$,
\begin{equation}\label{eitt}
\frac{b^{''}}{b}+\frac{1}{r}\frac{b^{'}}{b}-\left(\frac{b^{'}}{b}\right)^2
=-2b\left\{\Lambda+\frac{2\kappa^2{\cal T}_3
V}{\sqrt{X}}\left[\left(1+\frac{\tau^2}{br^2}\left(n-2C_{\theta}\right)^2\right)
\left(1+\frac{\tau^{'2}}{b}\right)+\frac{C_{\theta}^{'2}}{b^2
r^2}\right]\right\},
\end{equation}
\begin{eqnarray}\label{Neq}
\frac{N^{''}}{N}+\frac{1}{r}\frac{N^{'}}{N} =
-b\left\{2\Lambda+\frac{2\kappa^2 {\cal T}_3
V}{\sqrt{X}}\left[\left(1-\frac{E_{z}^2}{N^2}\right)
\left(2+\frac{\tau^{'2}}{b}+\frac{\tau^2}{br^2}\left(n-2C_{\theta}\right)^2
\right)-\frac{E_{r}^2}{N^2b}\right]\right\}.
\end{eqnarray}

Since the analytic structure of the system is complicated as shown
in (\ref{ter})--(\ref{Neq}), we begin with a simple but nontrivial
case. Specifically, the global D-strings with electric flux are
achieved by turning off the gauge field $C_{\theta}$ and the
electric field $E_{z}$. A choice for the cosmological constant is
$\Lambda=0$ since the minimum of the tachyon potential is naturally
set to be $V(\tau=\infty)=0$. Then the equations of motion for our
consideration are the tachyon equation (\ref{ter}) and the Einstein
equations (\ref{eitt})--(\ref{Neq}).

At the origin $(r=0)$, the boundary conditions are
\begin{equation}\label{bd0}
\tau(0)=0\quad (n\ne 0), \qquad N(0)=N_{0}, \qquad b(0)=b_{0},
\end{equation}
where $N_0$ and $b_0$ can be set to be unity by a rescaling.
Expansion of the fields near the origin gives
\begin{eqnarray}
\frac{\tau(r)}{\tau_0}&\approx &  r+\frac{Q ^2}{12} \left[
b_0\left(\kappa^4+(n^2-1)\frac{8{\cal T}_3^2}{Q^4} \right) + 2\tau
_0^2\left(\kappa^4+(n^2-1)\frac{4{\cal T}_3^2}{Q^4}\right) \right]
r^3 + \cdot\cdot\cdot ,
\label{t0}\\
\frac{N(r)}{N_{0}}&\approx &  1-\frac{|Q| \kappa^2 b_0}{\sqrt{ b_0 +
\tau_0 ^2}}\;r +
\frac{1}{2}Q^2 \kappa^4 b_0 \;r^2 + \cdot\cdot\cdot ,
\label{N0}\\
\frac{b(r)}{b_{0}}&\approx & 1-2|Q| \kappa^2\sqrt{b_0 + \tau_0 ^2}
\;r + \frac{1}{2} Q^2 \kappa^4 \left(5b_0 + 4 \tau_0^2 \right)r^2
+ \cdot\cdot\cdot ,
\label{b0}
\end{eqnarray}
where $\tau_0$ is the only undetermined parameter. The tachyon field
(\ref{t0}) is increasing from zero. The lapse function $N^{2}$
(\ref{N0}) and the conformal factor $b$ in front of the spatial part
(\ref{b0}) are decreasing from constant values.

\begin{figure}[ht]
\begin{center}
\scalebox{1.0}[1.0]{\includegraphics{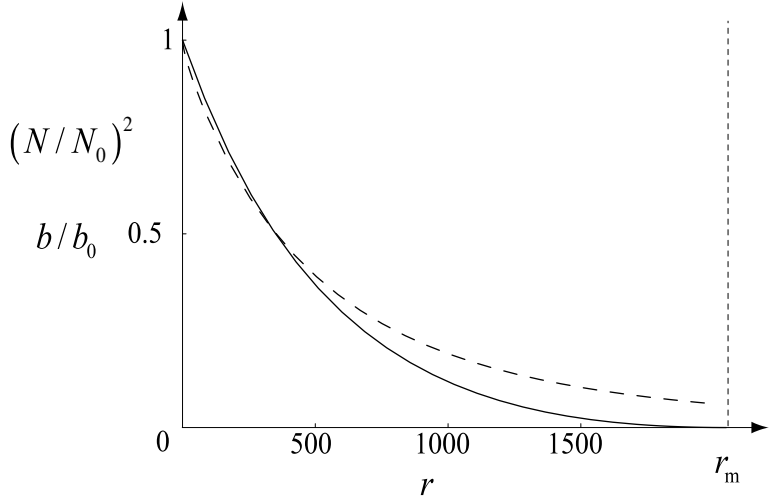}}
\par
\vskip-2.0cm{}
\end{center}
\caption{\small Graphs of the lapse function $N(r)^2$ (solid line)
and the conformal factor $b(r)$ (dashed line). Both of them are
monotonically decreasing in the range of $0\leq r < r_{\rm m}$.}
\label{fig1}
\end{figure}

\begin{figure}[ht]
\begin{center}
\scalebox{0.9}[0.9]{\includegraphics{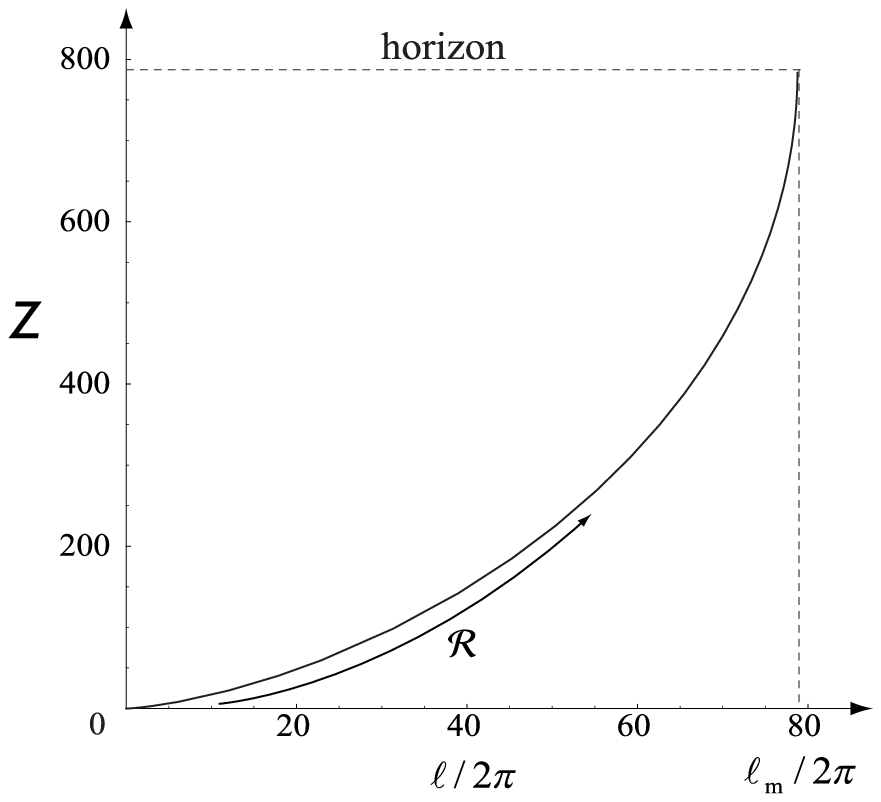}
\includegraphics{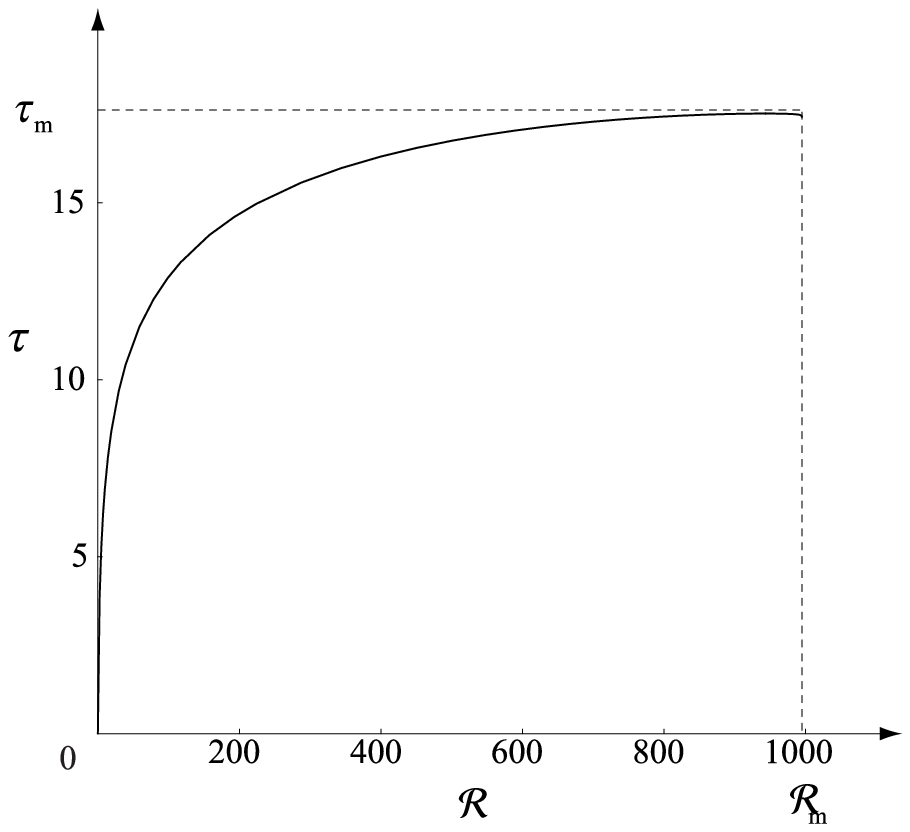}
}
\par
\vskip-2.0cm{}
\end{center}
\caption{\small The left graph shows that the geometry embedded in
three dimensions is a hemisphere and the right graph shows the
tachyon amplitude $\tau(r)$. The vertical axis ${\sf Z}$ in the left
graph is given by the relation, $d{\sf Z}^{2}=d{\cal
R}^{2}-d(\ell/2\pi)^{2}$ with ${\sf Z}(r=0)=0$, $\ell(r_{\rm
m})=\ell_{\rm m}$, and ${\cal R}(r_{\rm m})={\cal R}_{\rm m}$.}
\label{fig2}
\end{figure}

If we examine the tachyon equation (\ref{ter}) at a region of
sufficiently large $r$, we easily conclude that the infinite tachyon
amplitude at $r=\infty$ is impossible. It implies nonexistence of
the D-string solution monotonically connecting the false symmetric
vacuum, $\tau (0)=0$, and the true broken vacuum of the tachyon
potential (\ref{V3}), $\tau(\infty)= \infty$, in curved spacetime.
It is different from the case of flat spacetime. In this case, the
tachyon amplitudes of D- and DF-strings reach an infinite value at
the spatial infinity~\cite{Kim:2005tw,Cho:2005wp}. Therefor e, a
natural assumption is that its maximum value exists at a finite
radial coordinate $r_{{\rm m}}$ such that $\tau(r_{{\rm
m}})=\tau_{{\rm m}}$. A power series expansion supports this
assumption as
\begin{eqnarray}
\tau(r) &\approx & \tau_{{\rm m}} -\frac{b_{\rm m}\frac{4 {\cal T}_3
^2}{Q^2}e^{\frac{2\tau_{\rm m}}{R}}\left[ \left( e^{\frac{2\tau_{\rm
m}}{R}} -1 \right) \left(b_{\rm m} r_{\rm m} ^2 + n^2 \tau_{\rm m}^2
\right)\right] -n^2 R \tau_{\rm
m}}{R\left(1+e^{\frac{2\tau_{\rm m}}{R}}\right)\alpha}(r-r_{\rm m})^2 +\cdot\cdot\cdot,\label{tm}\\
\frac{N(r)}{N_{{\rm m}}} &\approx &  \left( r-r_{\rm m} \right)
-\frac{1}{2} \frac{N_{{\rm m}}^2}{r_{\rm m}} \left(r-r_{\rm
m}\right)^2 +
\cdot\cdot\cdot,\label{Nm}\\
\frac{b(r)}{b_{{\rm m}}} &\approx & 1 + b_{{\rm m}1} \left( r-r_{\rm
m} \right) +\left[ \frac{b_{{\rm m}1}^2}{2} -\frac{b_{{\rm
m}1}}{2r_{\rm m}} -\frac{|Q|\kappa^2  \sqrt{b_{{\rm m}}\alpha
}}{r_{\rm m} \left(1 +e^{\frac{2\tau_{{\rm m}}}{R}}
\right)}\right]\hspace{-1mm}\left( r-r_{\rm m}\right)^2 +
\cdots.\label{bm}
\end{eqnarray}
where\\
\begin{equation}
\alpha=1 +e^{\frac{4\tau_{{\rm m}}}{R}}+e^{\frac{2\tau_{{\rm
m}}}{R}}\left( 2+ \frac{16b_{{\rm m}}{\cal T}_3 ^2 r_{\rm m}^2}{Q^2
} + \frac{16 {\cal T}_3^2}{Q^2}n^2 \tau_{\rm m}^2 \right).
\end{equation}
The lapse function $N^{2}$ (\ref{Nm}) hits zero at $r_{\rm m}$ and
the time interval of a static observer to reach $r_{\rm m}$ is
diverging, $\int dt \sim \int d\tau  \  1/N_{{\rm m}}(r-r_{\rm
m})\stackrel{r\rightarrow r_{\rm m}}{\longrightarrow}\infty$. This
means that an event horizon is formed on the cylinder of radius
$r_{\rm m}$ parallel to the $z$-axis and the obtained global
($C_{\theta}=0$) D-string ($E_z=0$) with electric flux ($ E_r\neq0$)
is a black cosmic string~\cite{Kim:1997ba}. Numerical analysis shows
that it is indeed the case (See the solid line in Fig.~\ref{fig1}).
Since the conformal factor $b$ is decreasing near $r=r_{\rm m}$, the
signature of $b_{{\rm m}1}$ in (\ref{bm}) is negative, which is
supported by numerical analysis. Although it is decreasing  to a
nonzero value at $r_{{\rm m}}$, the radial distance from the origin,
${\cal R}(r)=\int_{0}^{r}dr^{'}\, \sqrt{b(r^{'})}\,$, is finite even
at the horizon $r_{{\rm m}}$. Correspondingly the circumference
along the angular coordinate $\theta$, $\ell(r)=r\sqrt{b(r)}$\,~, is
also finite. Therefore, the geometry between $r=0$ and $r=r_{{\rm
m}}$ is a hemisphere (See the left graph of Fig.~\ref{fig2}), and,
on this, the tachyon amplitude $\tau(r)$ grows from zero to the
maximum value $\tau_{{\rm m}}$ as given in ~(\ref{tm}) (See also the
right graph of Fig.~\ref{fig2}).

\begin{figure}[ht]
\begin{center}
\scalebox{1.0}[1.0]{\includegraphics{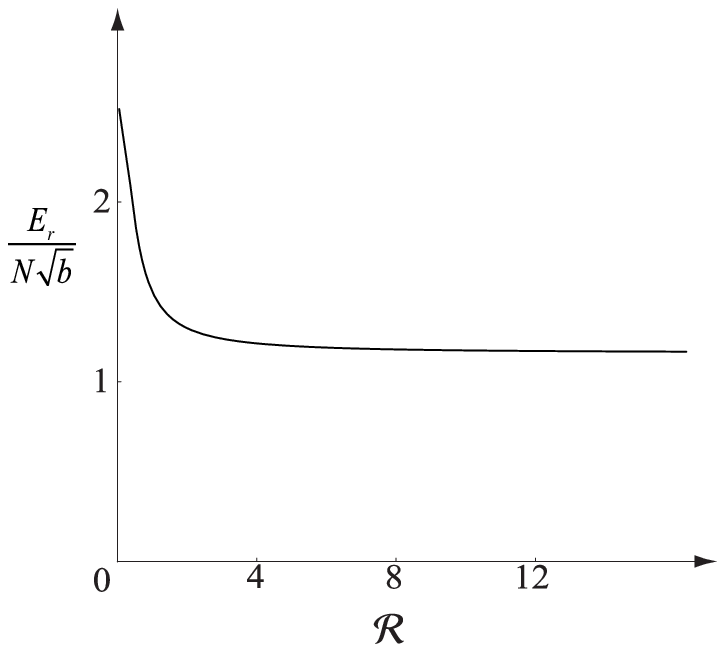}
\includegraphics{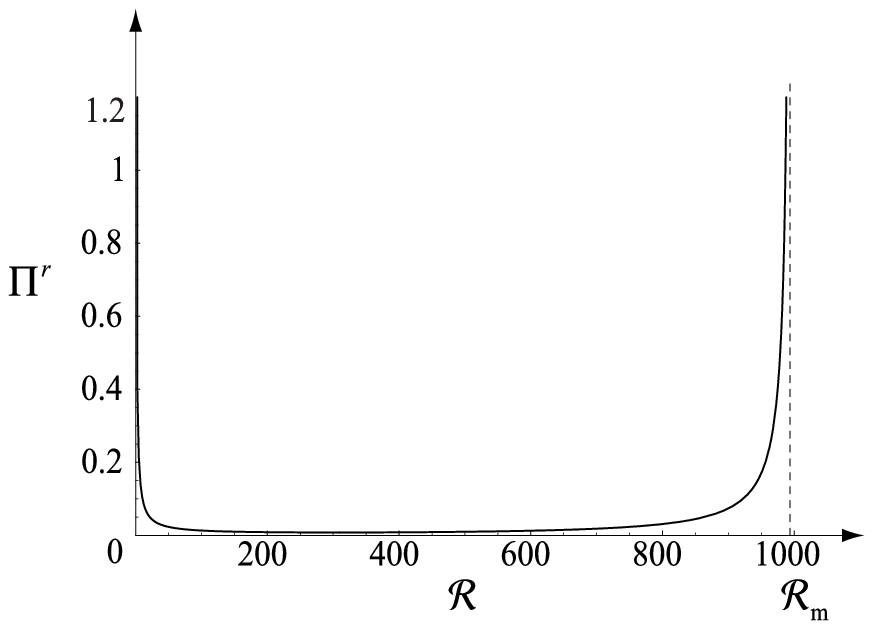}
}
\par
\vskip-2.0cm{}
\end{center}
\caption{\small Graphs of  the electric field
$\frac{E_{r}}{N\sqrt{b}}$ (left) and the radial component of
 the conjugate momentum $\Pi^{r}$ (right) .} \label{fig3}
\end{figure}

\begin{figure}[ht]
\begin{center}
\scalebox{1.0}[1.0]{\includegraphics{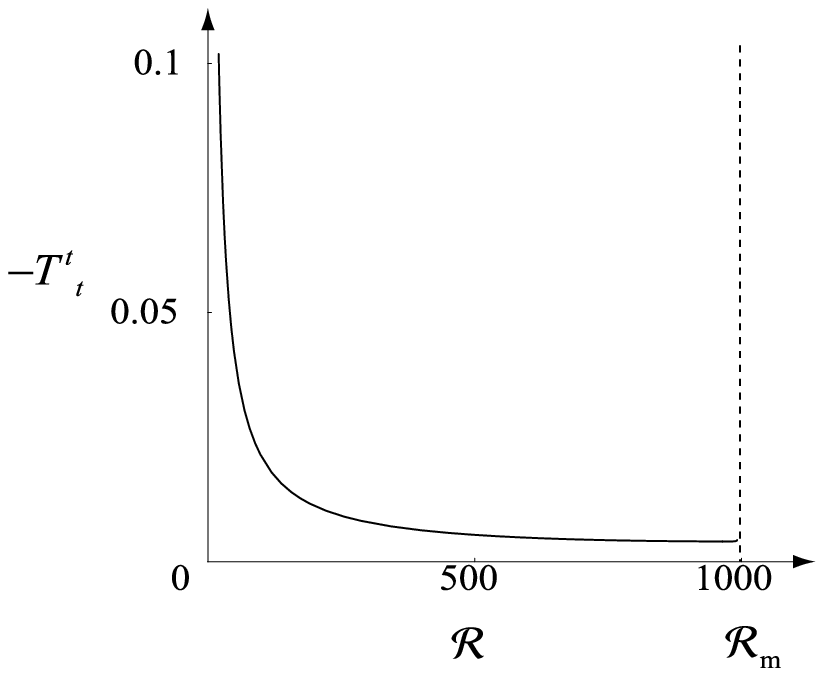}
\includegraphics{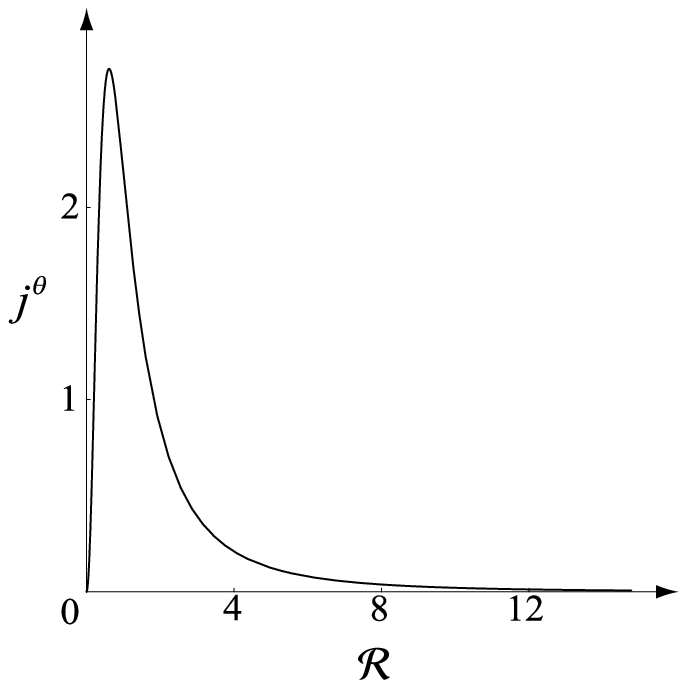}
}
\par
\vskip-2.0cm{}
\end{center}
\caption{\small Graphs of the energy density $-T^{t}_{\; t}$(left)
and $U(1)$ current $j^{\theta}$(right)} \label{fig4}
\end{figure}

To understand local physical properties of the straight global D-strings
with electric flux, we look into,
in addition to the radial component of electric field $E_{r}$ (\ref{er}),
nonvanishing components of the energy-momentum tensor
\begin{eqnarray}
-T^{t}_{\; t}&=&\frac{2{\cal T}_3
V}{\sqrt{X}}\left[\left(1+\frac{\tau^2}{br^2}\left(n-2C_{\theta}\right)^2\right)
\left(1+\frac{\tau^{'2}}{b}\right)+\frac{C_{\theta}^{'2}}{b^2
r^2}\right],
\label{emtt}\\
-T^{z}_{\; z}&=&\frac{2{\cal T}_3
V}{\sqrt{X}}\left[\left(1+\frac{\tau^2}{br^2}\left(n-2C_{\theta}\right)^2\right)
\left(1+\frac{\tau^{'2}}{b}-\frac{E_{r}^2}{N^2b}\right)
+\frac{C_{\theta}^{'2}}{b^2 r^2}\right],
\label{emzz}\\
-T^{r}_{\; r}&=&\frac{2{\cal T}_3
V}{\sqrt{X}}\left[1+\frac{\tau^2}{br^2}
\left(n-2C_{\theta}\right)^2\right]\left(1-\frac{E_{z}^2}{N^2}\right),
\label{emrr}\\
-T^{\theta}_{\; \theta}&=&\frac{2{\cal T}_3
V}{\sqrt{X}}\left[\left(1+\frac{\tau^{'2}}{b}\right)
\left(1-\frac{E_{z}^2}{N^2}\right)-\frac{E_{r}^2}{N^2b}\right],
\label{emhh}
\end{eqnarray}
and U(1) current $j^{\theta}$ for the gauge field $C_{\theta}$,
\begin{eqnarray}\label{jth}
j^{\theta}=\frac{2{\cal T}_3V}{\sqrt{X}}\left[
\left(1-\frac{E_{z}^2}{N^2}\right)
\left(1+\frac{\tau^{'2}}{b}\right)-\frac{E_{r}^2}{N^2
b}\right]\frac{\tau^2(n-2C_\theta)}{br^{2}}.
\end{eqnarray}
Insertion of the expansion of the fields near the origin
(\ref{t0})--(\ref{b0}) into (\ref{er}) and (\ref{emtt})--(\ref{jth})
leads to
\begin{eqnarray}
\frac{E_{r}}{N \sqrt{b}}&\approx& \sqrt{1 + \frac{\tau_0 ^2}{b_0}} +
\kappa^2|Q| \frac{\tau_0^2}{\sqrt{b_0}}\;r +{\cal O}(r^2),
\label{er0}\\
T^{t}_{\; t}&\approx& -\frac{|Q|\sqrt{b_0 +
\tau_{0}^2}}{b_0}\;\frac{1}{r} - \frac{Q^2 \kappa^2 (b_0 + 2\tau_0
^2)}{b_0}\nonumber\\
&&\hspace{-4mm} - \frac{|Q|^3 \sqrt{b_0 + \tau_{0}^2}}{16b_0}\left[
\frac{32{\cal T}_3^2}{Q^4}(b_0 +(2n^2-1)\tau_0^2)+4\kappa^4\left(b_0
+10\tau_0^2 \right)\right]r +{\cal
O}(r^2),\label{emtt0}\\
T^{z}_{\; z}&\approx&-\frac{4{\cal T}_3^2}{|Q|}\left(1+\frac{n^2
\tau_0^2}{b_0}\right)\sqrt{b_0+\tau_0^2} \;r+{\cal
O}(r^2),\label{emzz0}\\
T^{r}_{\;r}&\approx&-\frac{|Q|}{\sqrt{b_0+\tau_0^2}}\;
\frac{1}{r}-\frac{\kappa^2 Q^2 b_0}{b_0+\tau_0^2}
-\frac{|Q|^3}{16\left(b_0+\tau_0^2
\right)^{3/2}}\nonumber\\
& &\times \left[\frac{32{\cal T}_3^2}{Q^4}(b_0
+\tau_0^2)^2+4\kappa^4\left(b_0^2 -5b_0\tau_0^2 -2\tau_0^4\right)
\right]r+{\cal
O}(r^2),\label{emrr0}\\
T^{\theta}_{\; \theta}&\approx&-\frac{4{\cal
T}_3^2}{|Q|}\sqrt{b_0+\tau_0^2}\;r+{\cal
O}(r^2),\label{emhh0}\\
j^{\theta}&\approx&\frac{4n{\cal
T}_3^2\tau_0^2\sqrt{b_0+\tau_0^2}}{|Q|b_0}\;r + {\cal
O}(r^2).\label{jth0}
\end{eqnarray}
Insertion of the power series expansion near the horizon at $r_{{\rm
m}}$ (\ref{tm})--(\ref{bm}) into (\ref{er}) and
(\ref{emtt})--(\ref{jth}) provides
\begin{eqnarray}
\frac{E_{r}}{N \sqrt{b}}&\approx& \sqrt{\frac{1}{1+\frac{4{\cal
T}_3^2}{Q^2} \beta {\rm sech}\frac{\tau_{\rm m}}{R}}}
\left[1-\frac{\frac{r_{\rm m}}{b_{\rm m}} \left(2+b_{\rm m1}r_{\rm
m}\right)}{\left(\frac{Q^2}{2{\cal T}_3^2}\cosh\frac{\tau_{\rm
m}}{R}+2\beta \right)\cosh\frac{\tau_{\rm
m}}{R}}\left(r-r_{\rm m}\right)+\cdots\right],\label{erm}\\
T^{t}_{\; t}&\approx& T^{r}_{\;r} \nonumber\\ &\approx&-\frac{|Q|
\rm sech \frac{\tau_{\rm m}}{R}}{\sqrt{2b_{\rm m}}\;r_{\rm m}}
\sqrt{\gamma} \left[1-\frac{\left(2+b_{{\rm m}1} r_{\rm m}
\right)\delta}{2r_{\rm m}\gamma}\left(r-r_{\rm m}\right)+\cdots\right],\label{emtrm}\\
T^{z}_{\; z}&\approx&-\frac{4\sqrt{2}{\cal
T}_3^2\beta}{|Q|\sqrt{b_{\rm m}r_{\rm m}^2}\gamma}
\Bigg\{1+\frac{(2b_{\rm m}+b_{{\rm m}1} r_{\rm m})}{(b_{\rm m}r_{\rm
m}^2
+n^2\tau_{\rm m}^2)}\nonumber \\
& & \times\frac{\left[b_{\rm m}r_{\rm m}^2\left(1-n^2\frac{8{\cal
T}_3^2}{Q^2}+\cosh\frac{2\tau_{\rm m}}{R}\right) -n^2 \tau_{\rm
m}^2\delta\right]}{b_{\rm m}r_{\rm m}\gamma}
\left(r-r_{\rm m}\right)+\cdots\Bigg\},\label{emzzm}\\
T^{\theta}_{\; \theta}&\approx&-\frac{4\sqrt{2}{\cal
T}_3^2\sqrt{b_{\rm m}}r_{\rm m}{\rm sech} \frac{\tau_{\rm
m}}{R}}{|Q|\sqrt{\gamma }}\left[1+\frac{(2+b_{{\rm m}1}r_{\rm
m})\delta}{2r_{\rm
m}\gamma}\left(r-r_{\rm m}\right)+\cdots\right],\\
j^{\theta}&\approx&\frac{4\sqrt{2}n{\cal
T}_3^2\beta}{|Q|\sqrt{b_{\rm m}r_{\rm m}^2}\gamma }
\left[1-\frac{(2+b_{{\rm m}1}r_{\rm m})\left(\frac{8{\cal
T}_3^2}{Q^2}b_{\rm m}r_{\rm m}^2+\gamma\right)}{2r_{\rm
m}\gamma}\left(r-r_{\rm m}\right)+\cdots\right].\label{jthm}
\end{eqnarray}
where the parameters are
\begin{eqnarray}
\beta &=& \left(b_{\rm m}r_{\rm m}^2 +n^2\tau_{\rm m}^2\right)
{\rm sech}\frac{\tau_{\rm m}}{R},\\
\gamma&=& 1+\frac{8{\cal T}_3^2}{Q^2}(b_{\rm m}r_{\rm
m}^2+n^2\tau_{\rm m}^2)+\cosh \frac{2\tau_{\rm m}}{R},\\
\delta&=&1+ \frac{8{\cal T}_3^2}{Q^2}n^2\tau_{\rm m}^2 \cosh
\frac{2\tau_{\rm m}}{R}.
\end{eqnarray}
The radial component of electric field $\frac{E_{r}}{N\sqrt{b}}$ in
(\ref{erm}) is rapidly decreasing from a constant
 at the origin (\ref{er0}) to the other constant value at the horizon $r_{\rm m}$
(See the left graph in Fig.~\ref{fig3}). The fundamental string
charge density $\Pi^{r}(r)$ in (\ref{pir}) starts from an infinite
value at the origin, decreases to a finite value, and then diverges
to infinity at the horizon $r_{\rm m}$ (See the right graph in
Fig.~\ref{fig3}). Therefore, the fundamental strings come from a
transverse direction at the origin, flow along the latitude lines,
and go out to the transverse direction on the equator of the
hemisphere.
The energy density $-T^{t}_{\; t}$ and the radial component of
pressure $T^{r}_{\; r}$ are divergent at the origin as given in
(\ref{emtt0}) and (\ref{emrr0}) due to a contribution proportional
to $|Q|$ from the fundamental strings. They also approach
nonvanishing constants at the horizon as given in (\ref{emtrm}) (See
Fig.~\ref{fig4}).
For the pressure along the string direction $T^{z}_{\; z}$, the
angular components of pressure $T^{\theta}_{\; \theta}$, and the
$U(1)$ current $j^{\theta}$, they start from zero at the origin as
given in (\ref{emzz0}) and (\ref{emhh0})--(\ref{jth0}), vary
monotonically, and then reach constant values at the horizon as
given in (\ref{emzzm})--(\ref{jthm}). These nonvanishing constants
at the horizon imply the existence of a short hair. This seems to be
natural since the radial distance ${\cal R}$ from the location of
the global D-strings to the horizon is finite.

Let us discuss possible physical configurations outside the horizon
in what follows. At $r=\infty$, the following expressions can
provide a consistent set of solutions:
\begin{eqnarray}
&&\frac{\tau(r)}{\tau_\infty}\approx\frac{1}{r^{\frac{d}{2}-1}}\left[1+{\cal
O}\left(1/r^{\frac{3d}{2}-1}\right)\right],
\label{tf}\\
&&\frac{N(r)}{N_\infty}\approx 1-\frac{8b_\infty
\kappa^2|Q|}{(d-2)^2
\sqrt{4b_\infty+(d-2)^2\tau_\infty^2}}~\frac{1}{r^{\frac{d}{2}-1}}+{\cal
O}\left(1/r^{\frac{3d}{2}-1}\right),
\label{Nf}\\
&&\frac{b(r)}{b_\infty}\approx\frac{1}{r^d}
\left[1-\frac{4\kappa^2|Q|\sqrt{4b_\infty
+(d-2)^2\tau_\infty^2}}{(d-2)^2}~\frac{1}{r^{\frac{d}{2}-1}}+{\cal
O}\left(1/r^{\frac{3d}{2}-1}\right)\right]. \label{bf}
\end{eqnarray}
A (in fact unique) natural boundary condition at $r=\infty$ is the
vanishing tachyon amplitude and thus, (\ref{tf}) requires $d>2$.
Subsequently, as $ r\rightarrow\infty$, the circumference $\ell(r)$
along the angular coordinate $\theta$ goes to zero and the radial
distance from the origin ${\cal R}(r)$ is finite. Since the
asymptotic cylinder with finite $\ell(r=\infty)$ is excluded, a
candidate of such a two-dimensional spatial compact manifold can
uniquely be a two-sphere with $d=4$. Since this two-sphere is
produced by the vortices at the origin (we call $r=0$ as the south
pole), the topology of the two-sphere requires the vortices at
infinity (we call $r=\infty$ as the north pole). In addition, a
reflection symmetry between the south pole at $r=0$ and the north
pole at $r=\infty$ should be assigned~\cite{Gott:1984ef}. Through a
coordinate redefinition near the north pole as $k/r\rightarrow r$,
we fix the undetermined constants at infinity as $N_{\infty}=N_{0}$,
$\tau_{\infty}=k\tau_0$, and $b_{\infty}=k^2b_0$. Therefore, the
equator on the two-sphere should be located at the horizon
$(r=r_{\rm m})$, where the lapse function $N^2$ vanishes. If the
location of the horizon coincides with that of the
equator~\cite{Kim:2004xk}, the circumference $\ell(r)$ reaches the
maximum value along the equator, i.e., the condition
$d\ell/dr|_{r=r_{{\rm m}}}=0$ fixes a coefficient as $b_{{\rm
m}1}=-b_{{\rm m}}/r_{{\rm m}}$ in the conformal factor (\ref{bm}).
Then, the number of undetermined parameters $(\tau_{{\rm m}},
N_{{\rm m}}, b_{{\rm m}})$ in (\ref{t0})--(\ref{b0}), near the
horizon, is three which is the same as that near the origin
$(\tau_{0},N_{0},b_{0})$ in (\ref{tm})--(\ref{bm}).

Substituting the asymptotic functional behavior
(\ref{tf})--(\ref{bf}) into the local physical quantities in
(\ref{er}) and (\ref{emtt})--(\ref{jth}), we have
\begin{eqnarray}
\frac{E_{r}}{N \sqrt{b}}&\approx& \frac{\epsilon}{2\sqrt{b_{\infty}}}
+\frac{\kappa^2|Q|\tau_\infty^2}{\sqrt{b_\infty}}\;r^{1-d/2}
+{\cal O}(1/r^{\frac{3d}{2} -1}),
\label{erf}\\
T^{t}_{\; t}&\approx& -\frac{|Q|\epsilon}{4b_\infty}\; r^{-1+d/2}
- \frac{Q^2\kappa^2}{(d-2)^2 b_{\infty}}\,\epsilon^{2}
+{\cal O}(1/r^{\frac{d}{2} -1}),\\
T^{z}_{\; z}&\approx& -\frac{2{\cal T}_3^2}{|Q|}
\left(1+\frac{n^2\tau_\infty ^2}{b_\infty}\right)
\,\epsilon\, r^{1-d/2} +{\cal O}(1/r^{\frac{3d}{2} -1}),\\
T^{r}_{\;r}&\approx&-\frac{2|Q|}{\epsilon}
r^{-1+d/2}-\frac{16\kappa^2Q^2b_\infty}{(d-2)^2
\epsilon^{2}}+{\cal O}(1/r^{\frac{d}{2} -1}),\\
T^{\theta}_{\;\theta}&\approx& -\frac{2{\cal T}_3^2}{|Q|}
\,\epsilon\, r^{1-d/2}+{\cal O}(1/r^{\frac{3d}{2} -1}),\\
j^{\theta}&\approx& \frac{2n{\cal T}_3^2}{|Q|}
\frac{\tau_\infty^2}{b_\infty}\,\epsilon \, r^{1-d/2}
+{\cal O}(1/r^{\frac{3d}{2}-1}) ,
\label{jthf}
\end{eqnarray}
where $\epsilon=\sqrt{4b_\infty +(d-2)^2 \tau_\infty^2}\, $.
According to flipping $k/r\rightarrow r$ with
$\tau_{0}=\tau_{\infty}/k$, $N_{0}=N_{\infty}$, and
$b_{0}=b_{\infty}/k^2$, the coefficients of the leading terms near
the north pole (\ref{erf})--(\ref{jthf}) coincide exactly with those
near the south pole in (\ref{er0})--(\ref{jth0}). Therefore, the
physical quantities and two-sphere itself preserve a symmetry under
the exchange of the southern and northern hemispheres.

\begin{figure}[ht]
\begin{center}
\scalebox{1.0}[1.0]{\includegraphics{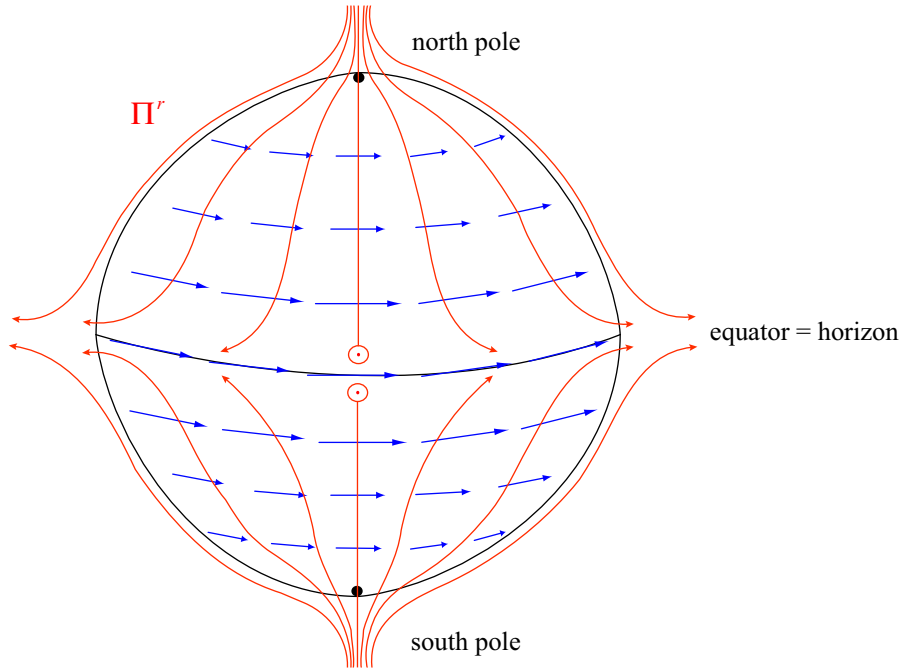}}
\par
\vskip-2.0cm{}
\end{center}
\caption{\small A schematic figure of global D-strings with
fundamental strings on ${\rm S}^{2}\times {\rm R}^{1}$. The
D-strings are on the south and north poles with the same vorticity.
Fundamental strings from a transverse direction come in at the south
and north poles and go out along the equator.} \label{fig5}
\end{figure}

The various aforementioned investigations suggest the following (See
also Fig.~\ref{fig5}). The spatial manifold of coordinates
$(r,\theta,z)$ is ${\rm S}^{2}\times {\rm R}^{1}$. There is a
horizon along the equator of the two-sphere. Global D-strings with
vorticity $n$, stretched along the $z$-axis $({\rm R}^{1})$, live at
both the south and north poles. Fundamental strings come from a
transverse direction to the south and north poles, flow to the
equator along the latitudes, and go out to the transverse direction
on the equator. Since the obtained horizon has a cylinder geometry,
${\rm S}^{1}\times {\rm R}^{1}$, surrounding each straight global
D-string at the south or north pole, the obtained D-string at one
hemisphere is identified as a straight black string to a static
observer at the other hemisphere. The resultant picture is
summarized in Fig.~\ref{fig5}.
Since the spatial geometry is ${\rm S}^{2}\times {\rm R}^{1}$ and
the field configuration keeps a translation symmetry along the
string direction of ${\rm R}^{1}$, the dimensionally-reduced
configuration is automatically constructed. To be specific, the
system of our interest is D2${\bar {\rm D}}2$, and we obtain
D-vortices  (or D0-branes) with a radial electric field, which are
located at both the north and south poles of ${\rm S}^{2}$. Since
the horizon on the equator (${\rm S}^{1}$) divides two D-vortices,
they are identified as two (1+2)-dimensional black holes.

\setcounter{equation}{0}
\section{BPS Limit of D- and DF-strings}
In this section, we discuss the systematic derivation of the BPS
limit for D- ($E_{z}=0$) and DF-strings ($E_{z}\ne 0$) stretched
parallel to the $z$-direction. When we study the BPS limit of a
solitonic object, rotational symmetry is not a necessary condition
so that we do not assume it in the case of flat spacetime. We show
that as expected, the BPS bound is saturated in the limit of zero
thickness for global ($C_\mu = 0$) D- and DF-strings without the
radial component of electric field ($E_{r}=0$). In this section, we
discuss first the case of flat spacetime and then that of the curved
spacetime.

Once a BPS limit is saturated, a necessary condition is vanishing
stress components, $T^{i}_{\; j}$. A softer condition would be the
vanishing pressure ($P^{i}=T^{i}_{\; i}$) difference,
\begin{eqnarray}
0=T^{i}_{\; i}-T^{|\epsilon_{ij}|j}_{\; |\epsilon_{ik}|k}&=&
\frac{2{\cal T}_3 V\sqrt{1-E_{z}^{2}}}{\sqrt{X|_{F=0}}}
(\partial_{i}{\bar T}\partial_{i}T-\partial_{|\epsilon_{ij}|j}{\bar T}
\partial_{|\epsilon_{ik}|k}T)\\
&=&\frac{{\cal T}_3 V\sqrt{1-E_{z}^{2}}}{\sqrt{X|_{F=0}}}
\left[(\overline{\partial_{i}T+ i\epsilon_{ij}\partial_{j}T})
(\partial_{i}T- i\epsilon_{ik}\partial_{k}T) \right.\nonumber\\
&&\hspace{30mm}+\left. (\overline{\partial_{i}T-
i\epsilon_{ij}\partial_{j}T}) (\partial_{i}T+
i\epsilon_{ik}\partial_{k}T) \right], \label{Bij}
\end{eqnarray}
where $i,j = 1,2$ (or $x,y$), $ii$ in $T^{i}_{\; i}$ denotes the
$i$-th pressure component, and
$X|_{F=0}=1+S_{ii}-\frac{1}{2}A_{ij}^2$ with $S(A)_{ij}\equiv
\frac{1}{2}\left(\partial_i{\overline{T}}
\partial_jT\pm\partial_j{\overline{T}}\partial_iT\right)$.
The right-hand side of (\ref{Bij}) should vanish. Then, we possibly
identify the Cauchy-Riemann equation as the first-order Bogomolnyi
equation for the BPS bound of arbitrary parallel D(F)-strings with
upper signature (or anti-D(F)-strings with lower signature);
\begin{eqnarray}\label{Be1}
(\partial_{i}\pm i\epsilon_{ij}\partial_{j})T=0,\qquad (\,
\partial_i \ln \tau = \pm \epsilon_{ij}\partial_j \chi \,).
\end{eqnarray}
We can easily confirm that substitution of the Bogomolnyi equation
(\ref{Be1}) leads to vanishing of the off-diagonal stress component
$T^{i}_{\; j}\;(i\ne j)$ for both the strings and anti-strings;
\begin{eqnarray}
T^{i}_{\; j}&=&
\frac{{\cal T}_3 V\sqrt{1-E_{z}^{2}}}{\sqrt{X|_{F=0}}}
(\partial_{i}{\bar T}\partial_{j}T
+\partial_{j}{\bar T}\partial_{i}T)\\
&=&\frac{{\cal T}_3 V\sqrt{1-E_{z}^{2}}}{2\sqrt{X|_{F=0}}}
\left[(\overline{\partial_{i}T\pm i\epsilon_{ik}\partial_{k}T})
(\partial_{j}T\mp i\epsilon_{jl}\partial_{l}T)
\right. \nonumber\\
&&\hspace{29mm}\left.
+(\overline{\partial_{j}T\pm i\epsilon_{jk}
\partial_{k}T})
(\partial_{i}T\mp i\epsilon_{il}\partial_{l}T) \right]
\label{sl}\\
&\stackrel{(\ref{Be1})}{=}&0.
\end{eqnarray}
Here, in the second line (\ref{sl}), we used $\epsilon_{ik}\partial_{k}=
\pm\partial_{j}$ and $\epsilon_{jk}\partial_{k}=\mp\partial_{i}$ for
$i\ne j$.

For the $n$ straight strings spread arbitrarily on the
$(x,y)$-plane, which are parallel along the $z$-direction, the phase
of the tachyon field becomes
\begin{eqnarray}\label{ph}
\chi= \pm\sum_{p=1}^{n}\tan^{-1}\frac{y-y_p}{x-x_p}.
\end{eqnarray}
Then, the tachyon amplitude $\tau$ is obtained as an exact solution
of the Bogomolnyi equation (\ref{Be1}),
\begin{eqnarray}\label{ta1}
\tau=\prod_{p=1}^{n}\tau_{{\rm BPS}} |{\bf x}-{\bf x}_p|,
\end{eqnarray}
where $\tau_{{\rm BPS}}$ is an integration constant and ${\bf
x}-{\bf x}_p=(x-x_{p}, y-y_{p})$. Due to the Bogomolnyi equation
(\ref{ta1}), we have
\begin{eqnarray}
S_{xx}&=&S_{yy}=\mp iA_{xy}\nonumber\\
&\hspace{-3mm}=&\hspace{-5mm}\tau_{{\rm
BPS}}^{2}\prod_{p=1}^{n}(\tau_{{\rm BPS}}|{\bf x}-{\bf x}_p|)^2
\sum_{q,r=1}^n\frac{\cos \theta_{qr}} {(\tau_{{\rm BPS}}|{\bf
x}-{\bf x}_q|)\; (\tau_{{\rm BPS}}|{\bf x}-{\bf x}_r|)}, \label{sxx}
\end{eqnarray}
where $\theta_{qr}$ is the angle between the two vectors, $({\bf
x}-{\bf x}_q)$ and $({\bf x}-{\bf x}_r)$. Plugging
(\ref{ph})--(\ref{sxx}) into the pressure components provides
$-T^{x}_{\; x}=-T^{y}_{\; y}=2{\cal T}_3 V\sqrt{1-E_z^2}\,$. Only in
the limit of zero thickness with $\tau_{{\rm BPS}}\rightarrow
\infty$, the pressure components vanish everywhere except the string
positions ${\bf x}_{p}=(x_{p},y_{p})$, i.e., $ S_{xx}|_{{\bf x}={\bf
x}_q}=0$ at ${\bf x}_p={\bf x}_q \;(p\neq q)$ and $-T^{x}_{\;
x}|_{{\bf x}={\bf x}_{p}}=-T^{y}_{\; y}|_{{\bf x}={\bf x}_{p}}
=2{\cal T}_3\sqrt{1-E_z^2}\,$.  Since (\ref{Be1}) gives
$X=X|_{F=0}\left(1-E_z^2\right)=(1-E_z^2)(1+S_{xx})^2$ and
$\partial_iT-A_{ij}\partial_jT=\partial_iT(1+S_{xx})$, the left-hand
side of the tachyon equation,
$\partial_i\left[\frac{V}{2\sqrt{X}}(1-E_z^2)(\partial_iT-A_{ij}\partial_jT)
\right]=\sqrt{X}\partial V/\partial\bar{T}$,  becomes
$\sqrt{1-E_z^2}\;S_{xx}\partial V/\partial\bar{T}$. Moreover,
$\partial V/\partial\bar{T }$ vanishes everywhere in the limit of
$\tau_{{\rm BPS}}\rightarrow \infty$. Thus the first-order
Bogomolnyi equation (\ref{Be1}) is consistent with the second-order
tachyon equation (\ref{ter}) for static BPS D- and DF-strings,
equivalent to conservation of energy-momentum,
$\partial_{j}T^{ji}=0$, for nontrivial solutions. Therefore, the
tachyon configurations given by the solutions
(\ref{ph})--(\ref{ta1}) of the Bogomolnyi equation (\ref{Be1}) can
be a BPS configuration only in the limit of infinite $\tau_{{\rm
BPS}}$.

To see the BPS relation, let us look into the Hamiltonian density:
\begin{eqnarray}\label{Ha}
{\cal H}=\sqrt{\Pi_{z}^2+4\Pi_{\bar{T}}\Pi_{T}+\left(
\Pi_{\bar{T}}\partial_i{\bar T}+\Pi_{T}\partial_i T\right)^2 +4{\cal
T}_3^2V^2 X|_{F=0}}\,\, ,
\end{eqnarray}
where the corresponding conjugate momenta are
\begin{eqnarray}
\Pi_{\bar{T}}&=&\frac{{\cal T}_3V}{\sqrt{X}}\left(\partial_{0}T
-A_{0i}\partial_iT\right),
\label{CM1}\\
\Pi_{T}&=&\frac{{\cal T}_3V}{\sqrt{X}}\left(\partial_{0}\bar{T}
+A_{0i}\partial_i\bar{T}\right),
\label{CM2}\\
\Pi_{z}&=&\frac{2{\cal T}_3V}{\sqrt{X}}E_{z} \left(X|_{F=0}\right).
\label{CM3}
\end{eqnarray}
For the static tachyon configurations, $\Pi_{{\bar T}}=\Pi_{T}=0$
from (\ref{CM1})--(\ref{CM2}). Then, the Hamiltonian density
(\ref{Ha}) reproduces the BPS relation for thin DF-strings
(($p,q$)-strings):
\begin{equation}\label{sHa}
{\cal H}=\sqrt{\Pi_{z}^2+\left(2{\cal T}_3V \sqrt{
X|_{F=0}}\right)^2}\,\, ,
\end{equation}
where the BPS formula for D-strings, ${\cal H}=|2{\cal T}_3V \sqrt{
X|_{F=0}}\,|$, is trivially obtained in the absence of the
fundamental string charge density $\Pi_{z} = 0$. Noticing that the
energy density (\ref{emtt}) is proportional to the electric flux
density $\Pi^{z}$ (\ref{piz}) as
\begin{equation}
{\cal H}=-\frac{\Pi^{z}}{E_{z}},
\end{equation}
we easily read for the DF-strings (($p,q$)-strings) that the charge
distribution of the fundamental string part coincides exactly with
that of the D-string part, which is confined at each string site in
the $(x,y)$-plane;
\begin{equation}
\Pi^{z}=-\frac{E_z}{\sqrt{1-E_z^2}} \;2{\cal T}_3V\sqrt{X|_{F=0}}\,.
\end{equation}

Inserting the BPS configuration (\ref{ph})--(\ref{ta1}) into the
energy density (\ref{ed1}), we arrive at
\begin{eqnarray} \label{BPe}
\int d^2 x \;{\cal H}= \frac{2{\cal T}_3}{\sqrt{1-E_z^2}} \int d^2x
\;\lim_{\tau_{{\rm BPS}}\rightarrow \infty} V\left(\tau \right)
\left(1+S_{xx}\right).
\end{eqnarray}
In the limit of infinite $\tau_{{\rm BPS}}$ with a rescaling to a
dimensionless variable $|{\tilde {\bf x}}-{\tilde {\bf
x}}_{p}|=\tau_{{\rm BPS}} |{\bf x}-{\bf x}_{p}|$, $S_{xx}$ is
proportional to $\tau_{{\rm BPS}}^{2}$ as given in (\ref{sxx}) and
is dominating in (\ref{BPe}) in comparison to unity. To obtain the
BPS sum rule and the corresponding descent relation for the
codimension-two D(F)-brane, ${\cal T}_1 = \pi^2 R^2 {\cal
T}_3/\sqrt{1-E_z^2}\,$, the required condition for $n$ D(F)-strings
is
\begin{eqnarray}
2{\cal T}_{1}|n|&=&  \frac{2{\cal T}_3}{\sqrt{1-E_z^2}}  \int d^2x
\; \lim_{\tau_{{\rm BPS}}\rightarrow \infty} V\left(\tau \right)
S_{xx}
\label{sxc}\\
&=&2\frac{\pi^{2}R^{2}{\cal T}_{3}}{\sqrt{1-E_{z}^{2}}}|n|.
\label{dsr}
\end{eqnarray}

Let us find a class of tachyon potentials saturating the BPS sum
rule from here on. As $\tau_{{\rm BPS}}$ approaches infinity for the
arbitrarily-distributed $n$ thin BPS D(F)-strings, $V(\tau)|_{{\bf
x} ={\bf x}_{p}}=1$ and then $V(\tau)S_{xx}\propto (\tau_{{\rm
BPS}})^{2n}$ diverges at each site of the strings ${\bf x}= {\bf
x}_{p}$. On the other hand, away from the string sites which have
${\bf x}\ne {\bf x}_{p}$ for all $p$, $V(\tau)=0$ and then
$V(\tau)S_{xx}$ vanishes. Therefore, the constraint to the shape of
the tachyon potential $V$ is
\begin{equation}
\lim_{\tau_{{\rm BPS}}\rightarrow \infty}V(\tau= \prod_{p=1}^{n}
\tau_{{\rm BPS}}|{\bf x}-{\bf x}_p|)S_{xx}=
\sum_{p=1}^{n}\delta^{(2)}({\bf x}-{\bf x}_{p}).
\end{equation}
Suppose that we consider first the 1/cosh type tachyon potential
(\ref{V3}) as a candidate. If we perform the integration for a
single D(F)-string with $S_{xx}=\tau_{{\rm BPS}}^{2}$ before taking
$\tau_{{\rm BPS}}\rightarrow \infty$, the integration produces
\begin{equation}
\int d^2(\tau_{{\rm BPS}}x)\;\frac{ 1}{\cosh
(\frac{\tau_{{\rm BPS}}|{\bf x}-{\bf x}_1|}{R})}=2\pi
K^2 R^2, \qquad \mbox{($K$ is the Catalan's constant)},
\end{equation}
which is different from the  descent relation (\ref{dsr}). This
result implies difficulty in employing this 1/cosh type tachyon
potential as a BPS potential for BPS D(F)-strings, which is
different from the case of tachyon kinks ~\cite{Kim:2003in}. Let us
take into account the Gaussian type tachyon potential obtained in
boundary string field theory~\cite{Gerasimov:2000zp}
\begin{equation}\label{bsft}
V(\tau)=\exp\left(-\frac{\tau^{2}}{\pi R^{2}}\right).
\end{equation}
Computation of the integration (\ref{sxc}) after substituting the
tachyon potential (\ref{bsft}) reproduces correctly the descent
relation (or the BPS sum rule) (\ref{dsr}) for $n$ superimposed
D(F)-strings with cylindrical symmetry, where we do not need to take
to the limit of infinite $\tau_{{\rm BPS}}$. Let us consider the $n$
separated D(F)-strings where the distance between any pair of two
strings is much larger than $1/\tau_{{\rm BPS}}$. In order to
perform the integration (\ref{sxc}), we look into a part of the
integrand $S_{xx}$ (\ref{sxx}) composed of a sum of the $n^{2}$
terms. For $n(n-1)$ terms with $q\ne r$, we show that the
integration of such terms becomes zero in the limit of infinite
$\tau_{{\rm BPS}}$. When ${\bf x}\ne {\bf x}_{s}~(s=1,2,...,n)$, the
integrand of every term vanishes by taking the limit due to
$\lim_{\tau_{{\rm BPS}\rightarrow \infty}}\exp(-{A}\tau_{{\rm
BPS}}^{\;\; 2n})B \tau_{{\rm BPS}}^{\;\; 2n}\rightarrow 0$ for any
nonvanishing finite $B$ and positive $A$. When ${\bf x}={\bf
x}_{s}$, $\tau$ in (\ref{ta1}) vanishes and $V(\tau)=1$ from
(\ref{bsft}). In addition, $S_{xx}\propto \left. |{\bf x}-{\bf
x}_{s}|\right|_{{\bf x}={\bf x}_{s}} \rightarrow 0$ when $s$ is
either $q$ or $r$ $(q\ne r)$, or $S_{xx}\propto \left. |{\bf x}-{\bf
x}_{s}|^{2}\right|_{{\bf x}={\bf x}_{s}} \rightarrow 0$ when $s$ is
neither $q$ nor $r$. Therefore, every $q\ne r$ term in (\ref{sxx})
does not contribute to the integrated BPS sum rule (\ref{dsr}) as
far as every pair of two D(F)-strings is separated with a distance
much larger than $1/\tau_{{\rm BPS}}$. If we finally take
$\tau_{{\rm BPS}}\rightarrow \infty$, the integration values of all
of the $q\ne r$ terms become zero irrespective of the nonvanishing
values of their separation distances. For the $n$ terms with $q=r$
in $S_{xx}$ (\ref{sxx}), the same argument as the case of $q\ne r$
is applied to the integrand in (\ref{sxc}) at every ${\bf x}$ away
from each string site, ${\bf x}\ne {\bf x}_{s}~(s=1,2,...,n)$. Now
the source of the nonvanishing integration value for (\ref{sxc})
comes from the $n$ terms of $q=r$ in $S_{xx}$ (\ref{sxc}) and the
contribution near each string site with $|{\bf x}-{\bf x}_{s}|\ll
|{\bf x}_{q}-{\bf x}_{s}|, \;\; (q=r\ne s)$. Near ${\bf x}={\bf
x}_{s}$, $\tau\rightarrow 0$ in (\ref{ta1}), $V(\tau)\rightarrow 1$
from (\ref{bsft}), and
\begin{equation}\label{apr}
S_{xx}\propto
\left\{
\begin{array}{ll}
\left. \cdots |{\bf x}-{\bf x}_{s}|^{2}\cdots
\right|_{{\bf x}={\bf x}_{s}}\rightarrow 0
& \mbox{for~} q=r\ne s \\
\displaystyle{ \lim_{\tau_{{\rm BPS}}\rightarrow \infty}
\prod_{{p=1}\;(p\ne s)}^{n} (\tau_{{\rm BPS}}|{\bf x}_{q}-{\bf
x}_{p}|)^{2} \rightarrow \infty } & \mbox{for~} q=r=s
\end{array}
\right. .
\end{equation}
We can perform this integration with sufficient accuracy since we
first take the limit of infinite $\tau_{{\rm BPS}}$ for the BPS
D(F)-strings. By using (\ref{apr}), in the limit of $\tau_{{\rm
BPS}}\rightarrow \infty$, the integration in (\ref{sxc}) is
rewritten as
\begin{eqnarray}
(\ref{sxc})&=& \frac{2{\cal T}_3}{\sqrt{1-E_z^2}}  \int d^2x \;
\lim_{\tau_{{\rm BPS}}\rightarrow \infty}
\sum_{s=1}^{n}\exp\left\{-\frac{1}{\pi R^{2}}\left[ \prod_{p=1\;
(p\ne s)}^{n}(\tau_{{\rm BPS}}|{\bf x}_{s}-{\bf x}_{p}|)^{2}
\right](\tau_{{\rm BPS}}|{\bf x}-{\bf x}_{s}|)^{2}\right\}
\nonumber\\
&&\hspace{48mm} \times \tau_{{\rm BPS}}^{\;\;\;\;\;\; 2} \left[
\prod_{q=1\; (q\ne s)}^{n}(\tau_{{\rm BPS}}|{\bf x}_{s}-{\bf
x}_{q}|)^{2} \right]. \label{sxd}
\end{eqnarray}
Since the integrand is a Gaussian type, this integration (\ref{sxd})
is easily performed in a closed form and results in the BPS sum rule
of our interest (\ref{dsr}). Moreover, we consider the case that a
part of the D(F)-strings are superimposed among the total $n$
D(F)-strings. Then, the above argument can be applied in the same
manner since the Gaussian type potential reproduces correctly the
BPS sum rule (\ref{dsr}) for an arbitrary number of superimposed
D(F)-strings as mentioned previously. In synthesis, the
aforementioned analysis means that the Gaussian type tachyon
potential (\ref{bsft}) passes all of the requirements for the BPS
sum rule (\ref{sxc})--(\ref{dsr}) so that (\ref{bsft}) is a BPS
potential for parallel D(F)-strings.

For these D- and DF-strings, the boost symmetry along the string
direction is sustained:
\begin{eqnarray}
-T^{t}_{\;t}=-T^{z}_{\;z} =\frac{2{\cal T}_3 V}{\sqrt{1-E_z^2}}
\left(1+S_{xx}\right). \label{ed1}
\end{eqnarray}
Though the above BPS limit was attained in the scheme of field
theory, the BPS limit for thin D(F)-strings is consistent with the
nature of a BPS D1-brane with zero thickness in string theory.

From now on, let us move to a discussion of the BPS bound in curved
spacetime. With the metric (\ref{trme}) for curved spacetime, the
pressure difference, $T^{r}_{\; r}-T^{\theta}_{\;\theta}$, becomes
\begin{eqnarray}\label{rme}
0=T^{r}_{\; r}-T^{\theta}_{\; \theta}
= \frac{2{\cal T}_3
V}{\sqrt{\left(1+\frac{\tau^{'2}}{b}\right)\left(1+n^2\frac{\tau^2
}{r^2 b}\right)}}\left(\tau^{'}
-n\frac{\tau}{r}\right)\left(\tau^{'}+n\frac{\tau}{r}\right)
\end{eqnarray}
which leads to the same BPS equation in flat spacetime (\ref{Be1})
with cylindrical symmetry and the linear ($|n|=1$) singular
($\tau_{{\rm BPS}}\rightarrow\infty$) tachyon profile
(\ref{ph})--(\ref{ta1}). We check that the first-order equation and
its would-be BPS configuration satisfies the second-order
Euler-Lagrange equations. Since the time- and angular components of
the energy-momentum conservation law is satisfied trivially
$\nabla_\mu T^{\mu}_{\; t}=\nabla_\mu T^{\mu}_{\; \theta}=0$, the
remaining radial component, $\nabla_\mu T^{\mu}_{\; r}=0$, becomes
equivalent to the tachyon equation,
\begin{eqnarray} \label{req}
\frac{d}{dr}\left[\frac{2{\cal
T}_3V}{\sqrt{X}}\left(1+\frac{\tau^2n^2}{br^2}\right)\right]
+\frac{2{\cal
T}_3V}{\sqrt{X}}\frac{N'}{N}\left(1+\frac{\tau^2n^2}{br^2}\right)
\frac{\tau^{'2}}{b^2}
-\left(\frac{1}{r}+\frac{b'}{2b}\right)(T^{r}_{\; r}-T^{\theta}_{\;\theta})
=0,
\end{eqnarray}
where $X=(1+\frac{\tau^2n^2}{br^2})(1+\frac{\tau^{'2}}{b})$. Note
that the Bogomolnyi equation $\tau'=\tau n/r$ from (\ref{rme}) makes
the equation (\ref{req}) much simpler with $(1+\frac{\tau^2
n^2}{br^2})/\sqrt{X}=1$. Its third term vanishes by (\ref{rme}) and
the linear singular solution making $V=0$ for $r\neq 0$ still could
be a consistent solution to (\ref{req}) only if the boundary
condition,
\begin{eqnarray}\label{bdy}
N'|_{r=0}=0,
\end{eqnarray}
is satisfied. In fact, $N(r)$ and $b(r)$ should be determined by
considering the Einstein equations (\ref{eitt})--(\ref{Neq}).
Substitution of (\ref{ta1}) with $|n|=1$ and $\tau_{{\rm
BPS}}=\infty$ results in
\begin{eqnarray}
&&\frac{(rN')'}{rN}=b\kappa^2\left(T^r_r+T^\theta_\theta\right)
=-2b\kappa^2{\cal
T}_3V\left(\sqrt{\frac{1+\frac{\tau^2n^2}{br^2}}{1+\frac{\tau^{'2}}{b}}}
+\sqrt{\frac{1+\frac{\tau^{'2}}{b}}{1+\frac{\tau^2n^2}{br^2}}}\, \right)
=-4b\kappa^2{\cal T}_3V , \label{Eeq2}
\\
&&\frac{1}{2rb}\left(\frac{rb'}{b}\right)'=\kappa^2T^t_t
=-2\kappa^2{\cal T}_3V\sqrt{\left(1+\frac{\tau^2n^2}{br^2}\right)
\left(1+\frac{\tau^{'2}}{b}\right)}
=-\frac{\delta(r)}{br}\pi\kappa^2R^2{\cal T}_3 , \label{Eeq1}
\end{eqnarray}
where the function $\delta(r)$ is defined as $\int_0^\infty
dr\delta(r)=1$. For $r\neq 0$, the right-hand sides of
(\ref{Eeq2})--(\ref{Eeq1}) vanish due to the runaway property of the
tachyon potential. Hence, the solutions for $r\neq 0$ are
$N=N_0+N_l{\rm ln}r$ and $b=b_0r^{q}$, where $N_0$, $N_l$, $b_0$,
and $q$ are integral constants. The nontrivial solution consistent
with the required boundary condition in (\ref{bdy}) is $N_{l}=0$.
Then (\ref{Eeq2}) requires a boundary condition
\begin{eqnarray}
br|_{r=0}=0,
\end{eqnarray}
since $V(r=0)$ is unity while $N'$ is zero in the entire region.
Hence, $q$ in $b=b_0r^q$ should be greater than $-1$.

In (\ref{Eeq1}), a constraint form of the tachyon potential for the
desired descent relation is assumed as discussed in (\ref{BPe}),
$\int_0^{2\pi}d\theta\int_0^\infty dr\sqrt{g_2}(-T^t_{\;
t})=2\pi^2R^2{\cal T}_3$, where $\sqrt{g_2}=br$ in this case. The
left-hand side of (\ref{Eeq1}) is
$\frac{1}{2\sqrt{g_2}}\partial_r(\sqrt{g_2}\partial^r{\rm
ln}b)=\frac{q}{2}\nabla^2{\rm ln}r=\frac{q}{2br}\delta(r)$.
Comparison with the right-hand side of (\ref{Eeq1}) yields
$q=-2\pi\kappa^2R^2{\cal T}_3$, which provides a familiar conic
spatial structure of the deficit angle,
$\Delta=2\pi(1-|1+q/2|)=-q\pi$. Since $q > -1$, the deficit angle
has, amazingly, the maximum value, $\pi$. Therefore, for the D- and
DF-strings, the same BPS limit is saturated in curved spacetime,
where the left-hand side of (\ref{Eeq1}) leads to a topological
quantity from the geometry side, an Euler number, and the right-hand
side to a topological charge of a soliton, vorticity.

In curved spacetime, comparing with the D-strings, the only
difference for the DF-string is a constant,
$\sqrt{1-E_{z}^{2}/N^{2}}\,$. Since BPS bound is attained for the
constant lapse function $N^{2}=N_{0}^{2}$, the previous BPS bound of
the D-strings in curved spacetime is automatically applied to the
case of DF-strings.

When the electric flux $E_{r}$ along the radial direction is turned
on, any BPS bound is not saturated for these thick D-strings with
electric flux. In addition, $-T^{t}_{\; t}\ne -T^{z}_{\; z}$ and
thus the boost symmetry along the string direction breaks down to a
translation symmetry along the $z$-axis.

For the Abelian Higgs model, the gauge field is indispensable. It
converts the global vortices of the logarithmically divergent energy
into the local vortices of finite energy. In addition, these local
vortices can saturate the BPS bound under a specific form of scalar
potential~\cite{MS}. In the case of D- and DF-strings, their energy
cannot be tamed by the gauge field $C_{\mu}$ and the BPS limit is
achieved in the limit of thin vortices without the gauge field
$C_{\mu}=0$. If we add the U(1) gauge field $C_{\mu}$ to make the
global strings be local ones and introduce the electric field
$E_{r}$ transverse to the strings, from
Eqs.~(\ref{emtt})--(\ref{emhh}) we could expect such additions to
hinder saturation of the BPS limit.

\setcounter{equation}{0}
\section{Conclusion and Discussion}

We studied gravitating global D- and DF-strings in a D3${\bar {\rm
D}}3$ system in the context of DBI type action of a complex tachyon
and a DBI electromagnetic field. Einstein gravity with a vanishing
cosmological constant has been assumed and only static straight
strings have been considered.

When the electric flux in the transverse direction is turned on, the
gravitated straight D-strings stretched along $z$-direction form a
black string. The resultant space is the product of a two-sphere and
a straight line, ${\rm S}^{2}\times {\rm R}^1$. The D-strings with
the same vorticity lie on both the north and south poles of the
two-sphere. The black string horizon is located on the equator so
that the geometry of the horizon is S${}^{1}\times$R$^{1}$. From a
transverse direction, fundamental strings also come into both the
north and south poles, flow along the latitudes, and go out through
the equator to the transverse direction. Although we constructed a
static black string solution, their stability~\cite{Gregory:1993vy}
needs to be addressed for the obtained structure of the black
strings (or branes)~\cite{Horowitz:1991cd,Kim:1997ba,Lee:2006jx}.

We demonstrated explicitly and systematically the BPS limit for thin
D- and DF-strings in both flat and curved spacetime. Almost all of
the BPS properties are shared with Nielsen-Olesen vortices in the
Abelian-Higgs model and their cosmic analogues, but the pressure
components on the $(x,y)$-plane have a nonvanishing value, $2{\cal
T}_{1}$, only at each D(F)-string site. We showed that a Gaussian
type tachyon potential satisfied the BPS sum rule consistent with
the descent relation for codimension-two BPS D(F)-strings. Our
language describing cosmic D- and DF-strings was DBI-type
action~\cite{Sen:2003tm,Garousi:2004rd}. It would be more intriguing
to describe the strings in the context of boundary conformal field
theory~\cite{Majumder:2000tt} or boundary string field
theory~\cite{Jones:2002si} as has been done for tachyon
kinks~\cite{Kim:2006mg}. In the scheme of the DBI-type field theory
of D3${\bar {\rm D}}3$ of our consideration, higher-derivative terms
have been neglected. Thus, their effect may be tested to check
whether or not the obtained BPS property is a genuine property of
BPS D1- or D1F1-branes in the string theory.

Once the BPS limit is systematically obtained, dynamics of moduli
space for the thin BPS D- and DF-strings~\cite{Copeland:2006eh}
should be tackled, which has been expected to be different from that
of the BPS Nielsen-Olesen vortices~\cite{Shellard:1988zx}. Then,
their cosmological implication is expected to lead to interesting
astrophysical phenomena in the early universe filled with cosmic
superstrings. In addition to the study of the static objects and
moduli space dynamics, full time evolution of D- and DF-strings by
solving numerically the classical equations of motion needs further
study~\cite{Felder:2002sv}.

\section*{Acknowledgments}
We would like to thank Inyong Cho, Chanju Kim, Sung-Won Kim, and
Kyungha Ryu for helpful discussions. This work is the result of
research activities (Astrophysical Research Center for the Structure
and Evolution of the Cosmos (ARCSEC)) and grant No.\
R01-2006-000-10965-0 from the Basic Research Program supported by
KOSEF (T.K. and Y.K.), and by the Korea Research Foundation Grant
funded by the Korean Government (MOEHRD) (KRF-2006-312-C00095)
(J.L.).

\end{document}